# Oscillation growth in mixed traffic flow of human driven vehicles and automated vehicles: Experimental study and simulation

Shi-Teng Zheng[1], Rui Jiang[1,*], H.M. Zhang[2,*], Junfang Tian[3], Ruidong Yan[1], Bin Jia[1], Ziyou Gao[1]

[1] *Key Laboratory of Transport Industry of Big Data Application Technologies for Comprehensive Transport, Ministry of Transport, Beijing Jiaotong University, Beijing 100044, China*
[2] *Department of Civil and Environmental Engineering, University of California Davis, Davis, CA 95616, United States*
[3] *Institute of Systems Engineering, College of Management and Economics, Tianjin University, Tianjin 300072, China*

## ABSTRACT

This paper reports an experimental study on oscillation growth in mixed traffic flow of automated vehicles (AVs) and human driven vehicles (HVs). The leading vehicle moves with constant speed in the experiment. The following vehicles consist of six developable AVs and different number of HVs. Thus, the market penetration rate (MPR) of AVs decreases with the increase of platoon size. The AVs are homogeneously distributed in the platoon. The constant time gap car-following policy is adopted for the AVs and the gap is set to 1.5 s. The experiment shows that in the 7-vehicle-platoon, the oscillations grow only slightly. In the 10-vehicle-platoon, the AVs could still significantly suppress the growth of oscillations. With the further decrease of MPR of AVs in the 13- and 20-vehicle-platoon, the AVs become having no significant impact on oscillation growth. On the other hand, with the decrease of MPR of AVs, average density of the vehicles and flow rate of the platoon increase, which demonstrates a trade-off between traffic stability and throughput under the given setup of AVs. The simulation study is also carried out, which exhibits good agreement with the experiment. Finally, sensitivity analysis of the parameters in the AV upper-level control algorithm has been performed, which is expected to guide future experiment design.

*Keywords:* Car following; Field experiment; Mixed traffic flow; Automated vehicle; Oscillation growth

## 1. Introduction

Automated vehicles (AVs) are expected to have profound impacts on traffic system. In recent years, the market penetration rate (MPR) of commercial AVs is increasing rapidly, which is predicted over 10% in 2025 (Calvert et al., 2017). With the rapid increase of MPR of AVs, traffic flow characteristics would be fundamentally changed in the near future (Auld et al., 2017; Soteropoulos et al., 2019).

The main benefits of emerging AV technologies can be summarized as follows: (i) dampening traffic oscillation and stabilizing traffic flow (see, e.g., Shang and Stern, 2021; Stern et al., 2018); (ii) improving throughput and avoiding breakdown (see, e.g., Kesting et al., 2008; Talebpour and Mahmassani, 2016); (iii) saving energy and mitigating pollution (see, e.g., Ard et al., 2020; Zhang and Ioannou, 2006); and (iv) enhancing driving safety (see, e.g., Chirachavala and Yoo, 1994; Kikuchi et al., 2003).

However, recently, some empirical evidences of commercial AVs revealed that the commercial AVs are string unstable (see, e.g., Gunter et al., 2021; Gunter et al., 2020; Makridis et al., 2020b). This important finding indicates that the impact of AV technologies on the traffic flow might be negative. Therefore, one should not take AVs as a sure mean of increasing throughput or stability for granted.

Although AV technology is developing at an unprecedented speed, its industrialization and popularization may take decades. Therefore, the mixed traffic situations that consist of both AVs and human driven vehicles (HVs) correspond to likely deployment scenarios in the long transition towards 100% deployment (Mahmassani, 2016).

---

* Corresponding authors.
E-mail addresses: jiangrui@bjtu.edu.cn (R. Jiang), hmzhang@ucdavis.edu (H.M. Zhang).

In the past several decades, many efforts have already been devoted to the impact of AVs in the mixed traffic flow of AVs and HVs. These studies mainly focus on the traffic stability, capacity, and energy saving in the mixed traffic flow. For details, one may refer to literature review in Section 2.

Recently, a series of experimental studies of car following behavior of HVs demonstrate that traffic oscillation grows in a concave way along the platoon (Jiang et al., 2014, 2015, 2018). The finding reveals the mechanism of traffic instability and exhibits how traffic flow develops into jams; thus, it plays an important role in traffic dynamics, which would help to understand, model and manage traffic oscillations. However, the study on the impact of AVs on traffic oscillation growth is lacked.

To address this gap, we performed a large-scale field experiment of mixed platoon of AVs and HVs. Note that the upper-level control algorithm of the AVs used in our experiment is set by the users, and the lower-level controller can be separately studied from the upper-level controller. This enables the users to better understand the traffic dynamics of the AVs than that of the commercial AVs. Under different MPR of AVs, we report: (i) how the oscillation grows along the mixed traffic platoon, and how the average velocity and platoon length fluctuate; (ii) how the AVs impact the flow-density relation; and (iii) how to simulate the mixed traffic flow.

The rest of paper is organized as follows. In Section 2, a detailed literature review on the impact of AVs in the mixed traffic flow is presented. Section 3 provides the experimental setup. Section 4 introduces the AV controllers. Section 5 reports the experimental results. Section 6 presents the simulation study. Finally, Section 7 concludes the paper.

## 2. Literature review

Since the concept of intelligent cruise control (ICC) system was proposed in the late 1980s and early 1990s (see, e.g., Broucke and Varaiya, 1996; Minderhoud and Bovy, 1999; Watanbe et al., 1995; Zhang and Benz, 1993), many efforts have been devoted to the impact of AVs on traffic flow[1], which can be roughly classified into simulation based studies and field experiment based ones, as summarized in Tables 1-3.

### 2.1. Simulation based studies

As early as in Chirachavala and Yoo (1994), the potential impact of the ICC system on traffic operation safety was evaluated based on simulation. The results showed that the introduction of ICC system would not lead to traffic perturbation problems and could help to enhance driver comfort and safety. Minderhoud and Bovy (1999) performed the capacity analysis for an on-ramp bottleneck on a two-lane freeway. It was shown that the headway setting as well as the ICC design parameters, e.g., the speed range, has impact on traffic capacity. Marsden et al. (2001) carried out a simulation study towards the understanding of Adaptive Cruise Control (ACC) vehicles. They found that following an ACC vehicle can reduce the variation of acceleration. Thus, driving comfort and environmental benefits can be achieved. Kesting et al. (2008) utilized the intelligent driver model (IDM) to describe both HVs and ACC vehicles. Simulations revealed the improvement of traffic stability and capacity as well as travel times with the introduction of ACC vehicles. Spiliopoulou et al. (2018) also simulated both HVs and ACC vehicles using the IDM. They studied the potential benefits of ACC vehicles on the recurrent traffic congestion at an on-ramp bottleneck and found that the congestion can be reduced even at low MPRs. Goni-Ros et al. (2019) carried out microscopic simulation of a freeway with a sag bottleneck and showed that ACC vehicles are able to reduce traffic congestion at the sag under high-demand condition. Kim et al. (2021) modeled AVs based on the MIXIC car-following model and predicted that AVs can reduce traffic congestion.

---

[1] The ICC or Adaptive Cruise Control (ACC) vehicle realizes the longitudinal automation. Therefore, in terms of car-following behavior, the ICC or ACC vehicle can be treated as an AV.

Some works further considered the cooperative ACC (CACC) vehicles or connected AVs (CAVs), which can communicate with its neighboring vehicles or infrastructure via wireless communication. For example, Van Arem et al. (2006) used the MIXIC model to simulate the CACC vehicles and studied the impact of CACC vehicles on the traffic flow characteristics near a lane-drop bottleneck. It was found that CACC vehicles can improve the traffic stability and efficiency. Xiao et al. (2018) studied the influence of CACC on highway with merging bottlenecks, considering the CACC deactivation to ACC or HV mode, which leads to the flow heterogeneity. The relation between CACC deactivation, traffic congestion and flow heterogeneity was studied. A strong correlation between traffic congestion and number of CACC deactivations was found. Moreover, although the increase of MPR can increase the capacity, the capacity drop persists at all CACC MPRs. Liu et al. (2018) studied the impact of CACC vehicles on traffic capacity and quantified the improvement at a four-lane freeway corridor and on/off-ramp area. With the increase of MPR, a quadratic trend of the capacity was identified. Ramezani et al. (2018) simulated the freeway corridor mixed traffic with heavy trucks and passenger cars, in which they introduced the CACC trucks and studied the traffic improvement under different MPRs of CACC trucks in terms of vehicle miles traveled, average speed and flow rate. It was found that CACC trucks not only reduce congestion and improve the traffic operations, but also have no adverse effect on car operations. Talebpour and Mahmassani (2016) utilized the IDM to describe CAVs and assessed analytically the string stability of heterogeneous traffic under different MPRs. In addition to improve the traffic stability, they argued that the CAV technologies have the potential to improve the throughput by more than 100%.

Apart from microscopic simulations, there are also simulation works using macroscopic models. For instance, Ngoduy (2012) proposed an extended multiclass gas-kinetic theory for mixed traffic flow of HVs and ACC vehicles, and revealed that ACC vehicles contribute to the improved stabilization of traffic flow and about 30% MPR of ACC vehicles can significantly increase capacity and reduce travel time. Huang et al. (2020) modeled the mixed traffic of CAVs and HVs using continuum traffic flow models, and conducted the linear stability analysis to quantify the impact of CAVs on the stability. It was also found that increasing MPR and sensitive reaction of CAVs result in more significant stabilizing effect. Zheng et al. (2020) proposed a stochastic model for mixed traffic flow with HVs and AVs considering the uncertainty of human driving behavior and analyzed the impact of AVs on uncertainty and stability of mixed traffic flow. The simulation results showed that large MPR of AVs can reduce the uncertainty of traffic flow and thus substantially improve the stability of mixed flow. On the other hand, the reaction time of AVs and the position of AVs in the platoon have subtle impact on the uncertainty of mixed flow.

In contrast to the positive effect of AVs, some studies suggested that in the mixed traffic flow, the impact of AV technologies is not significant and may be negative in some certain scenarios. For instance, VanderWerf et al. (2001, 2002) defined a set of mathematical models and carried out the simulations to predict the effect of ACC vehicles on traffic dynamics and capacity. It was shown that ACC vehicles only have a small impact on highway capacity even under the most favorable conditions. Shladover et al. (2012) evaluated the highway capacity under varying MPRs of ACC vehicles. They found that the introduction of ACC vehicles is unlikely to change capacity significantly. Olia et al. (2018) also showed that the traffic capacity appears insensitive to the MPR of ACC vehicles. Jerath and Brennan (2012) modeled HVs and ACC vehicles using the General Motors (GM) car-following model with different model parameter values. It was demonstrated that the introduction of ACC vehicles might improve the traffic flow rate, or result in traffic congestion under different circumstances. Van Arem et al. (2006) pointed out that there is a degradation of performance for a dedicated lane of CACC vehicles when the MPR is lower than 40%. Liu et al. (2018) reported when the CACC MPR is larger than 60%, the bottleneck capacity decreases significantly with the off-ramp traffic percentage due to the difficulty for off-ramp vehicles to find acceptable gaps to make lane changes.

Table 1. Studies based on simulations.

| Literature | HV Model | AV model | Main results |
|---|---|---|---|
| Chirachavala and Yoo (1994) | GM model | Vehicle control laws | • The ICC system can reduce the frequencies of hard acceleration/deceleration, enhance speed harmonization in traffic and cut down the incidence of "less-safe" headway.<br>• The ICC system can potentially reduce traffic accidents by 7.5%. |
| Minderhoud and Bovy (1999) | SiMoNe (Minderhoud and Bovy, 1998) | SiMoNe | • Headway setting and the ICC design parameters, e.g., the speed range and reactivation method, have impact on roadway capacity.<br>• When the headway setting is 1.2 s, there is no significant change on roadway capacity near an on-ramp bottleneck regardless of the ICC type. |
| Marsden et al. (2001) | FLOWSIM model (Wu et al., 1998) | ACC algorithm based on a manufacturer prototype | There is considerable reduction in the variation of acceleration when following an ACC vehicle, which indicates potential comfort gain and environmental benefits. |
| Kesting et al. (2008) | IDM (Treiber et al., 2000) | IDM | • ACC vehicles can improve the traffic stability and capacity.<br>• Traffic congestions can be completely eliminated with 25% ACC vehicles.<br>• Travel times can be significantly reduced even for 5% ACC vehicles. |
| Spiliopoulou et al. (2018) | IDM | IDM | The space-time extent of congestion can be reduced even at low MPRs of ACC vehicles, which also results in the improvement of average vehicle delay and fuel consumption. |
| Goni-Ros et al. (2019) | Modified IDM (Schakel et al., 2010) | A non-linear state-feedback controller with constant time-gap policy | • The increasing ACC vehicles can reduce traffic congestion under high-demand condition, and no congestion at the sag for more than 75% MPR.<br>• Advanced ACC systems (e.g., CACC) perform better than the basic ACC in reducing congestion, which can prevent the formation of congestion at a lower MPR. |
| Kim et al. (2021) | In the environment of VISSIM | MIXIC car-following model (Van Arem et al., 1995) | • AVs can increase road capacity.<br>• With the increase of traffic demand, high MPRs of AVs present better performance. |
| Van Arem et al. (2006) | MIXIC car-following model (Van Arem et al., 1997) | MIXIC car-following model | • CACC vehicles contribute to an improvement of traffic stability and a slight increase in efficiency.<br>• There is a degradation of performance for a dedicated lane of CACC vehicles when the MPR is lower than 40%. |
| Xiao et al. (2018) | Modified IDM | Speed control and gap control laws | • The increase of CACC MPR can increase the traffic capacity.<br>• There is a strong correlation between traffic congestion and number of CACC deactivation.<br>• Capacity drop persists in bottleneck scenarios at all CACC MPRs.<br>• CACC vehicles increase flow heterogeneity due to the switch among different operation modes. |
| Liu et al. (2018) | NGSIM oversaturated flow model (Yeo et al., 2008) | Gap regulation controller with constant time gap policy (Milanés and Shladover, 2014) | • The capacity improvement has a quadratic trend with the CACC MPR.<br>• The improvement of throughput decreases with the increase of on-ramp traffic.<br>• When the CACC MPR is larger than 60%, the bottleneck capacity decreases significantly with the off-ramp traffic percentage due to the difficulty for off-ramp vehicles to find acceptable gaps to make lane changes. |
| Ramezani et al. (2018) | In the environment of AIMSUN using Micro-SDK | Speed control and gap control laws | • CACC trucks can reduce congestion propagation and improve traffic operations for trucks with respect to vehicle miles traveled, average speed and flow rate. |

| Literature | HV Model | AV model | Main results |
|---|---|---|---|
| | | | • CACC trucks do not adversely affect car operations and even result in a significant speed increase. |
| Talebpour and Mahmassani (2016) | An extension of acceleration model (Talebpour et al., 2011) | IDM | • CAVs can improve string stability.<br>• CAVs are more effective in preventing shockwave formation and propagation.<br>• Substantial throughput increases under certain penetration scenarios. |
| Ngoduy (2012) | Gas-kinetic model (Hoogendoorn and Bovy, 2000) | Gas-kinetic model | • ACC vehicles can improve the traffic stability.<br>• About 30% MPR of ACC vehicles can result in significantly increased capacity and reduced travel time. |
| Huang et al. (2020) | Aw-Rascle-Zhang model | Mean field game traffic flow model | • The increasing MPR of CAVs results in more significant stabilizing effect.<br>• The more sensitively CAVs react to HVs, the more stable the mixed traffic is. |
| Zheng et al. (2020) | Stochastic mixed traffic model (Zhen et al., 2018) | Stochastic mixed traffic model | • Large MPRs of AV can reduce the uncertainty and substantially improve the stability.<br>• The reaction time of AVs has subtle impact on the uncertainty of mixed flow.<br>• The position of AVs in the platoon has marginal influence in reducing uncertainty and improving stability. |
| VanderWerf et al. (2001, 2002) | Cognitive and control human driver model (Song and Delorme, 2000) | Linear ACC controller with constant time gap policy | • Even under the most favorable conditions, ACC vehicles only have a small impact on highway capacity.<br>• The capacity benefits decrease quickly with the increase of ACC MPR. In particular, when the MPR is above 60%, there is a loss of capacity. |
| Shladover et al. (2012) | NGSIM oversaturated flow model | Speed control and gap control laws | ACC vehicles are unlikely to change the capacity significantly. |
| Olia et al. (2018) | Fritzsche car-following model in Paramics | Speed control and gap control laws | The traffic capacity appears insensitive to the MPR of ACC vehicles. |
| Jerath and Brennan (2012) | GM model (Gazis et al., 1959) | GM model | The introduction of ACC vehicles may produce higher flow rate, or result in traffic congestion under different circumstances. |

## 2.2. Field experiment based studies

The simulations are usually based on simplifications and assumptions. For example, in quite a few AV models, the lower-level controller is not considered. In some other AV models, the simple first-order lag model is used for the lower-level controller. On the other hand, most studies used deterministic HV models, which is not able to reproduce the nontrivial stochastic human driving behavior. As a result, the studies may not conform to the real behaviors of mixed traffic flow of AVs and HVs. To address this gap, in recent years, there have been considerable efforts devoting to experimental studies.

Some studies used two or three vehicles in the experiments. For example, Ioannou and Stefanovic (2005) carried out a set of tests with three vehicles involving one ACC vehicle to examine the energy saving of ACC vehicle under the disturbances of high acceleration, lane cut-in and lane exiting. The results showed that the ACC vehicle can smooth these disturbances and is also beneficial to the environment when compared with the similar all HV situations. Shi et al. (2021) and Shi and Li (2021) studied the fuel consumption and traffic stability of commercial AVs by conducting the experiments with one HV and two AVs under different speed ranges and headway settings. They found that a larger headway results in more stable traffic flow and less fuel consumption, and AVs always require less fuel consumption than HVs. Nevertheless, larger headway setting may decrease the capacity compared with all HV traffic. Rahmati et al. (2019) conducted several car-following tests utilizing three-vehicle platoon with an AV in between, and evaluated the difference in the behavior of the rear HV. They showed that for human drivers, following an AV is more comfortable. Similarly, Zhao et al. (2020) designed a set of two-vehicle tests and studied the difference on longitudinal characteristics of the HV between following a HV and following an AV. It was found that the response of the following HV depends on the driver's subjective trust on AV, which suggests that the mixed traffic characteristic may be complex due to the characteristic of human drivers. Ding et al. (2021) used three Tesla vehicles with SAE level-2 autonomous driving function to carry out an experiment on an open road and captured the significant heterogeneity for different operating scenarios (i.e., AV-AV, AV-HV, HV-AV, HV-HV). Makridis et al. (2020a) tested two vehicles with a commercial ACC vehicle following a controlled HV to understand the operational domain of ACC controller with regard to the response time and the time gap. It was found that the ACC response time is comparable to human driver, while the time gap is a little higher than HVs. Gunter et al. (2020) conducted the experiments with a commercial ACC vehicle following a pre-specified vehicle and pointed out that the ACC system is string unstable. Li et al. (2021) adopted the platoon with two commercial ACC vehicles following a HV and studied the car-following characteristics of ACC vehicles in different conditions. The results showed that the behaviors of ACC vehicles are complex and depend on the ACC headway setting, speed level and leader stimulus, which underlines the importance of empirical test to understand such complexity.

There are also quite a few experimental studies that used longer platoon with more than three vehicles. For example, Milanés and Shladover (2014) conducted a four-vehicle platoon experiment and validated that the multiple consecutive ACC vehicles are unstable due to the large delay in the response of following vehicles. Ge et al. (2018) tested the connected cruise controller using CAV and connected HVs (CHVs) and demonstrated the benefits of V2X-based control by mitigating traffic waves. It was also shown that both safety and energy efficiency can be significantly improved for the CAV and its neighboring CHVs. Stern et al. (2018, 2019) conducted the experiment on a circular track, involving more than 20 vehicles with one Cognitive and Autonomous Test (CAT) vehicle, which focuses on the dampening effect of CAT vehicle on traffic waves and vehicle emissions. They argued that the traffic flow control will be possible with less than 5% mobile actuators, which can dampen traffic waves and largely reduce vehicle emissions. Knoop et al. (2019) tested a platoon that consists of seven ACC vehicles on public freeway, and found that the ACC system is unstable, which could lead to collision risks in oscillation situations when the platoon is longer than three or four vehicles. Also under the actual freeway conditions, He et al. (2020) tested a five-vehicle mixed platoon with two HVs for the leader and the last vehicle and three ACC vehicles in between. They found that the mixed platoon is unstable and there is a negative impact of ACC vehicles on energy consumption. Makridis et al. (2020b) tested a platoon of five ACC vehicles under different perturbation conditions and showed that the platoon is unstable in these situations. Gunter et al. (2021) also conducted an eight-AV platoon test to validate the string instability. Ciuffo et al. (2021) summarized the results of an 11-car platoon with 10 commercial ACC vehicles under different vehicle orders and headway settings. The results confirmed the string instability of ACC vehicles depends on time gap setting, and also showed that the string unstable behavior of ACC vehicles would lead to higher energy consumption and new safety risks.

## 3. Experimental setup

The experiment was performed on Oct. 12 and 13, 2021, on a about 1.5-kilometer straight track in the closed test field affiliated to Research Institute of Highway, Ministry of Transport, China. For the satellite picture of test track and the snapshot of experiment, see Fig. 1.

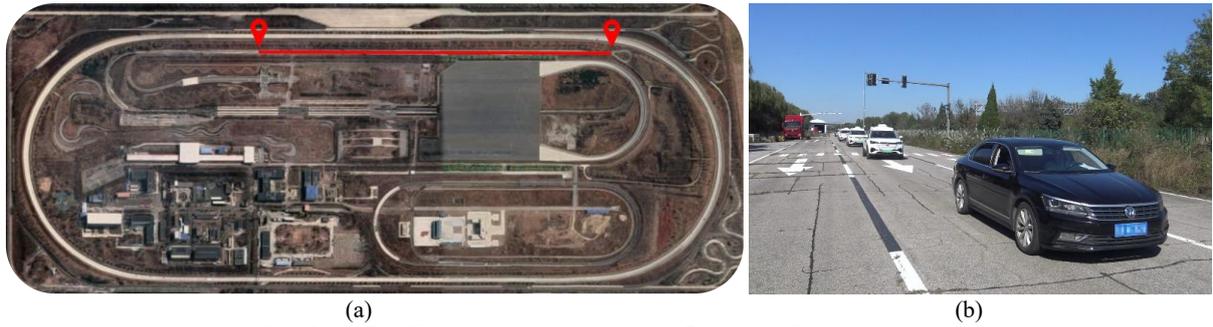

(a)                                                                                          (b)

Fig. 1. (a) Satellite picture of test track. (b) Snapshot of the experiment.

In the experiment, we employed six developable AVs. The algorithm and parameter values of upper-level controller of the AVs can be set by the users. The AVs can switch into HV mode as needed. The leading vehicle is equipped with cruise control (CC) system and thus can move with constant speed by switching on the CC system. Other HVs have no CC system and are always driven by human drivers. See Appendix A for the detailed information of experimental vehicles. During the two-day experiment, four different sized platoons are used, i.e., the 7-, 10-, 13-, 20-vehicle-platoon. In the platoons, the six AVs are homogenously distributed. Excluding the leading vehicle, the MPR of AV is 100%, 67%, 50%, 32%, respectively. To illustrate the impact of AVs in the mixed traffic flow, the benchmark experiment is carried out in the 10-, 13-, 20-vehicle-platoon scenario, in which all AVs switched into HV mode.

The order of vehicles/drivers in the four sized platoons is always fixed. The trajectories of vehicles were collected by AV itself and/or independent GPS devices. For the AVs, the equipped control computer calculates the speed difference, spacing and acceleration according to the information from lidar, GPS and other on-board vehicle sensors every 0.05 s. Moreover, high precision independent GPS devices have been installed at the identical left-rear position of all vehicles to record their locations and velocities every 0.1 s. The measurement errors of the GPS devices are within ±1 m for location and within ±1 km/h for velocity.

In the experiments, all vehicles moved straight ahead and did not change lane. The AV follows its preceding vehicle under the pre-specified upper-level control algorithm in the longitudinal direction, while the steering wheel of AVs is always controlled by human drivers. The instruction to the HV drivers and AV drivers (in HV mode) is, "Drive as if you were on rush hour expressways. Do not change lane to overtake. Follow the vehicle ahead closely, whenever safety permits".

Initially the vehicles are stopped bumper-to-bumper. When an experimental run starts, the leading vehicle accelerates to the given cruise speed $v_l$ and moves at that speed until the end of the experimental run. Once the last vehicle stops, the platoon will make a U-turn and prepare the next run. Settings of the experimental Runs are shown in Tables 4-7 for the 7-, 10-, 13-, 20-vehicle-platoon, respectively.

Table 2. Studies based on field experiments using two or three vehicles.

| Literature | Vehicle composition | Scenario | Main results |
|---|---|---|---|
| Ioannou and Stefanovic (2005) | Two HVs and one ACC vehicle | The leading HV executes a given speed profile, and the disturbance situations of high acceleration from the leading HV and lane cut-in and lane exiting of the other HV are tested. | The ACC vehicle response smooths the disturbance created by the high acceleration, cut-in or exiting of the vehicle, which is also beneficial to the environment compared with the similar all HV situations. |
| Shi et al. (2021); Shi and Li (2021) | One HV and two AVs | The leading vehicle executes the pre-specified speed profiles.<br>• CF experiment: For different operating scenarios, i.e., AV-AV, AV-HV, HV-AV, HV-HV, under different speed ranges and headway settings.<br>• Platoon experiment: HV-AV-AV | • With the AV headway setting increases, the AV string stability increases and the overall fuel consumption decreases.<br>• With the speed of AV traffic increases, the impact of AV headway settings on fuel consumption decreases.<br>• AVs always require less fuel consumption than HVs for the same experiment settings.<br>• The structure of triangular fundamental diagram is suitable for AV traffic.<br>• While the shortest AV headway setting can improve traffic capacity, larger headway setting may decrease capacity compared with all HV traffic. |
| Rahmati et al. (2019) | A three-vehicle mixed platoon with two HVs and one AV in between | The leading vehicle follows a series of given speed profiles. The second vehicle can be in AV mode or switch to HV mode. Data are collected from the third vehicle. | • The difference on the car-following behavior is statistically significant between the HV-AV and HV-HV operations for following HV.<br>• It is more comfortable for human driver following in the HV-AV operation. |
| Zhao et al. (2020) | One HV and one AV | The leading AV executes the constant or dynamic speeds; the HV follows the AV. | The response of following HV to the leading AV depends on driver's subjective trust on AV rather than the actual driving behavior. |
| Ding et al. (2021) | Three Tesla vehicles equipped with SAE level-2 automation | The AVs are asked to drive back and forth on the test segment of an open road, during which the interaction between AVs and HVs is captured by a camera on a high building. | There is significant heterogeneity on the car-following behavior between AVs and HVs in mixed traffic. |
| Makridis et al. (2020a) | Two consecutive vehicles, in which the following one is ACC vehicle | The vehicles are driven under different desired speeds and time gap settings. | • The response time of the tested ACC controller is in the range of 0.8-1.2 s, which is comparable to human drivers.<br>• The time gap of the ACC system is in the range of 1.2-2.2 s, which is slightly higher than the desired time gap of human drivers. |
| Gunter et al. (2020) | Two consecutive vehicles, in which the following one is ACC vehicle | The leading HV executes the given speed profiles, consisting of a variety of constant speeds and dynamic conditions where the speed is changed quickly. | • The tested ACC vehicle is string unstable.<br>• Commercial ACC platoon of moderate size can dampen some disturbances even it is string unstable. |
| Li et al. (2021) | Three-vehicle platoon with two following ACC vehicles | The leading vehicle executes different given speed profiles, including different traffic conditions and stimulus. | • The ACC response time is comparable to human drivers.<br>• The behaviors of ACC vehicles largely depend on the ACC headway setting, speed level and leader stimulus.<br>• The change from one ACC vehicle to the next is progressive for oscillation growth, and regressive for deceleration, acceleration and overshooting. |

Table 3. Studies based on field experiments using more than three vehicles.

| Literature | Vehicle composition | Scenario | Main results |
|---|---|---|---|
| Milanés and Shladover (2014) | Four consecutive ACC vehicles | The leading vehicle executes the given speed profile, consisting of a series of speed changes. | • The response of following vehicles has a large delay to cause the unstable response.<br>• String of consecutive ACC vehicles is unstable, which amplifies the speed variation of preceding vehicles. |
| Ge et al. (2018) | • Exp 1: A CAV travels on a road with a right turn behind a CHV.<br>• Exp 2: A CAV travels on a straight road behind three CHVs.<br>• Exp 3: A CAV travels on a straight road behind six CHVs and followed by a CHV. | • Exp 1: The leading HV is beyond the line of sight of a human driver or sensors due to road geometry.<br>• Exp 2: The leading vehicle decelerates and triggers many severe decelerations.<br>• Exp 3: The leading vehicle maintains the average speed of 20 m/s with mild braking events. | • V2X-based control can compensate for the beyond-line-of-sight failures due to road geometry and improve performance.<br>• The CAV can respond with additional "phase lead" and avoid the sudden speed changes with the help of motion information from CHV ahead.<br>• The CAV can mitigate the traffic waves by using information from multiple preceding CHVs.<br>• Both safety and energy efficiency can be significantly improved for the CAV and the neighboring CHVs. |
| Stern et al (2018, 2019) | 20-21 HVs together with a CAT vehicle on a single-lane ring track | The CAT vehicle is initially under human control, and the autonomous controller is activated when the traffic waves are noticeably observed. | • Traffic waves can be dampened by controlling the velocity of a single vehicle, and thus the flow control will be possible via less than 5% mobile actuators.<br>• Vehicle emission of the entire fleet may be reduced by 15%-73% when stop-and-go waves are dampened by CAT vehicle. |
| Knoop et al. (2019) | Seven vehicles equipped with SAE level-2 automation | The platoon travels on the public freeway driving at the speed similar to the surrounding traffic, during which different traffic conditions are encountered. | • It is difficult to create platoons intentionally on busy public roads.<br>• Since the platoon is unstable, it is not suitable for driving as platoon longer than three or four vehicles. |
| He et al. (2020) | A five-vehicle platoon with two HVs for the leader and the last vehicle and three ACC vehicles in between | The platoon travels at the freeway under actual traffic conditions. | • ACC followers lead to string instability.<br>• On the individual level, ACC followers have energy consumption 2.7-20.5% higher than those of all HV scenarios.<br>• On the platoon level, the energy values of ACC followers tend to consecutively increase 11.2-17.3%. |
| Makridis et al. (2020b) | A five ACC vehicle platoon | The leading vehicle executes a pre-specified speed profiles with different constant speeds and perturbation conditions. | • The platoon is string unstable.<br>• The response time of the controllers is in the range of 1.7-2.5 s, significantly longer than that reported in the literature.<br>• The range of time headway settings is quite broad. |
| Gunter et al. (2021) | An eight-vehicle platoon with seven following ACC vehicles | The leading vehicle executes a pre-specified speed profile, consisting of a speed slow down event. | Commercially implemented ACC systems are string unstable. |
| Ciuffo et al. (2021) | An 11-vehicle platoon with ten following ACC vehicles | The leading vehicle executes the pre-specified speed profiles, consisting of different equilibrium speeds of 30-60 km/h and several perturbations of different magnitudes. | • ACC vehicles are string unstable, which is not an intrinsic feature but depends on the algorithmic implementation by manufacturers.<br>• ACC vehicles will possibly lead to higher energy consumption and introduce new safety risks with the increase of MPR.<br>• ACC vehicles can activate traffic hysteresis in the presence of string unstable behavior.<br>• Road slope has a reinforcement effect on string instability. |

Table 4. Settings of the experimental Runs of 7-vehicle-platoon.

| Run No. | $v_l$ (km/h) | AV status | Platoon composition |
|---|---|---|---|
| 1 | 20 | AV mode | Leader-**AV**-**AV**-**AV**-**AV**-**AV**-**AV** |
| 2 | 30 | | |

Table 5. Settings of the experimental Runs of 10-vehicle-platoon.

| Scenario | Run No. | $v_l$ (km/h) | AV status | Platoon composition |
|---|---|---|---|---|
| Benchmark | 1 | 20 | HV mode | Leader-**AV**-**AV**-HV-**AV**-**AV**-HV-**AV**-**AV**-HV |
| | 2 | | | |
| | 3 | 30 | | |
| | 4 | | | |
| Mixed | 1 | 20 | AV mode | |
| | 2 | | | |
| | 3 | | | |
| | 4 | 30 | | |
| | 5 | | | |

Table 6. Settings of the experimental Runs of 13-vehicle-platoon.

| Scenario | Run No. | $v_l$ (km/h) | AV status | Platoon composition |
|---|---|---|---|---|
| Benchmark | 5 | 20 | HV mode | Leader-**AV**-HV-**AV**-HV-**AV**-HV-**AV**-HV-**AV**-HV |
| | 6 | | | |
| | 7 | 30 | | |
| | 8 | | | |
| Mixed | 1 | 20 | AV mode | |
| | 2 | | | |
| | 3 | 30 | | |
| | 4 | | | |
| | 5 | | | |

Table 7. Settings of the experimental Runs of 20-vehicle-platoon.

| Scenario | Run No. | $v_l$ (km/h) | AV status | Platoon composition |
|---|---|---|---|---|
| Benchmark | 9 | 20 | HV mode | Leader-HV-**AV**-HV-HV-**AV**-HV-HV-HV-**AV**-HV-HV-**AV**-HV-HV-HV-**AV**-HV-HV-HV |
| | 10 | | | |
| | 11 | 30 | | |
| | 12 | | | |
| Mixed | 1 | 20 | AV mode | |
| | 2 | | | |
| | 3 | | | |
| | 4 | 30 | | |
| | 5 | | | |
| | 6 | | | |

# 4. The controllers of AV

This section introduces the controllers implemented on the AVs in the experiment, which would help to understand the vehicle dynamics of the AVs.

## 4.1. Upper-level control algorithm

In our study, the constant time gap car-following policy is adopted in the upper-level longitudinal controller to generate the command acceleration, which reads

$$a_{n,cmd}(t) = k_g \cdot \left( \Delta x_n(t) - T_g v_n(t) - G_{\min} \right) + k_v \cdot \Delta v_n(t) \tag{1}$$

where $a_{n,cmd}$ is the command acceleration of vehicle $n$; $\Delta x_n$ is the measured spacing from rear axle of vehicle $n$ to the center of preceding vehicle $n-1$; $v_n$ and $\Delta v_n$ are the speed and the measured speed difference with the preceding vehicle, respectively; $k_g$ and $k_v$ are the control gains; $T_g$ is the constant time gap; $G_{\min}$ is the standstill spacing. In the experiment, the parameter values of upper-level controller are set to $k_v = 0.3$ s$^{-1}$, $k_g = 0.3$ s$^{-2}$, $T_g = 1.5$ s and $G_{\min} = 9.5$ m.

## 4.2. Lower-level controller

The AV executes the command from upper-level controller via the lower-level controller, i.e., vehicle mechanical structure. This process can be described in frequency domain as follows.

$$A(s) = G(s) A_{cmd}(s) \qquad (2)$$

where $A_{cmd}(s)$ and $A(s)$ denote the command acceleration and the actual one in frequency domain, respectively; $G(s)$ denotes the transfer function from $A_{cmd}(s)$ to $A(s)$.

While users can customize the upper-level control algorithm of the AVs, the lower-level controller is determined by vehicle mechanical structure, which is usually unknown to the users and may vary from vehicle to vehicle. For the modeling of lower-level controller, the most commonly used model is the first-order lag model (Liang and Peng, 1999; Wang, 2018; Wen et al., 2019), which reads

$$G_1(s) = \frac{1}{T_d s + 1} \qquad (3)$$

where $T_d$ is the time delay.

Another commonly used lower-level control model is the second-order response model (Flores and Milanés, 2018; Milanés et al., 2014), which reads

$$G_2(s) = \frac{k}{s^2 + 2\theta\omega s + \omega^2} e^{-T_d s} \qquad (4)$$

where $k$ denotes the static gain; $\theta$ and $\omega$ denote the damping factor and natural frequency, respectively.

### 4.3. Calibration and validation for lower-level control model

We test the performance of the two aforementioned lower-level control models based on the experimental data. They are calibrated using the genetic algorithm in the Optimization Toolbox of MATLAB with the goodness-of-fit function

$$\min \sum_{Run} \sqrt{\frac{1}{T_{Run}} \sum_{t=1}^{T_{Run}} \left( a_{t,Run}^{sim} - a_{t,Run}^{exp} \right)^2} \qquad (5)$$

where $a_{t,Run}^{sim}$ and $a_{t,Run}^{exp}$ denote the actual acceleration at time $t$ in the simulation and in the experimental Run, respectively; $T_{Run}$ denotes the time length of the experimental Run. We randomly choose 70% Runs used for calibration, and the rest of Runs are used for validation.

The calibrated model parameter values of lower-level control models $G_1(s)$ and $G_2(s)$ are presented in Tables 8 and 9, respectively. The comparison of actual acceleration profile between experiment and simulation are shown in Fig. 2. One can see that both models perform well. The average calibration errors are 0.2314 m/s² and 0.1932 m/s² for $G_1(s)$ and $G_2(s)$, respectively. The average validation errors are 0.2367 m/s² and 0.2073 m/s², respectively. As expected, the second-order response model slightly outperforms the first-order lag model. Therefore, the second-order response model is utilized in the simulation study in Section 6.

Table 8. Calibrated model parameters of lower-level controller $G_1(s)$.

| Parameter | Unit | Lower bound | Upper bound | Calibrated | | | | | |
|---|---|---|---|---|---|---|---|---|---|
| | | | | AV 1 | AV 2 | AV 3 | AV 4 | AV 5 | AV 6 |
| $T_d$ | s | 0 | 3 | 0.4622 | 0.4986 | 0.4583 | 0.9338 | 0.5718 | 0.4882 |
| *Calibration error* | m/s² | | | 0.1940 | 0.1842 | 0.1898 | 0.2914 | 0.3770 | 0.1522 |
| *Validation error* | m/s² | | | 0.2141 | 0.1979 | 0.1896 | 0.2859 | 0.3670 | 0.1655 |

Table 9. Calibrated model parameters of lower-level controller $G_2(s)$.

| Parameter | Unit | Lower bound | Upper bound | Calibrated | | | | | |
|---|---|---|---|---|---|---|---|---|---|
| | | | | AV 1 | AV 2 | AV 3 | AV 4 | AV 5 | AV 6 |
| $\theta$ | — | 0 | Inf | 0.4901 | 0.4985 | 0.4923 | 0.369 | 0.3753 | 0.6355 |
| $\omega$ | s⁻¹ | 0 | Inf | 4.4433 | 4.3659 | 4.2829 | 3.7857 | 3.7663 | 4.2634 |
| $k$ | s⁻² | 0 | Inf | 13.847 | 15.199 | 15.550 | 10.435 | 9.0481 | 16.042 |
| $T_d$ | s | 0 | 3 | 0.1993 | 0.2222 | 0.1878 | 0.6245 | 0.1817 | 0.1475 |
| *Calibration error* | m/s² | | | 0.1657 | 0.1673 | 0.1763 | 0.2390 | 0.2725 | 0.1382 |
| *Validation error* | m/s² | | | 0.1902 | 0.1815 | 0.1802 | 0.2504 | 0.2915 | 0.1500 |

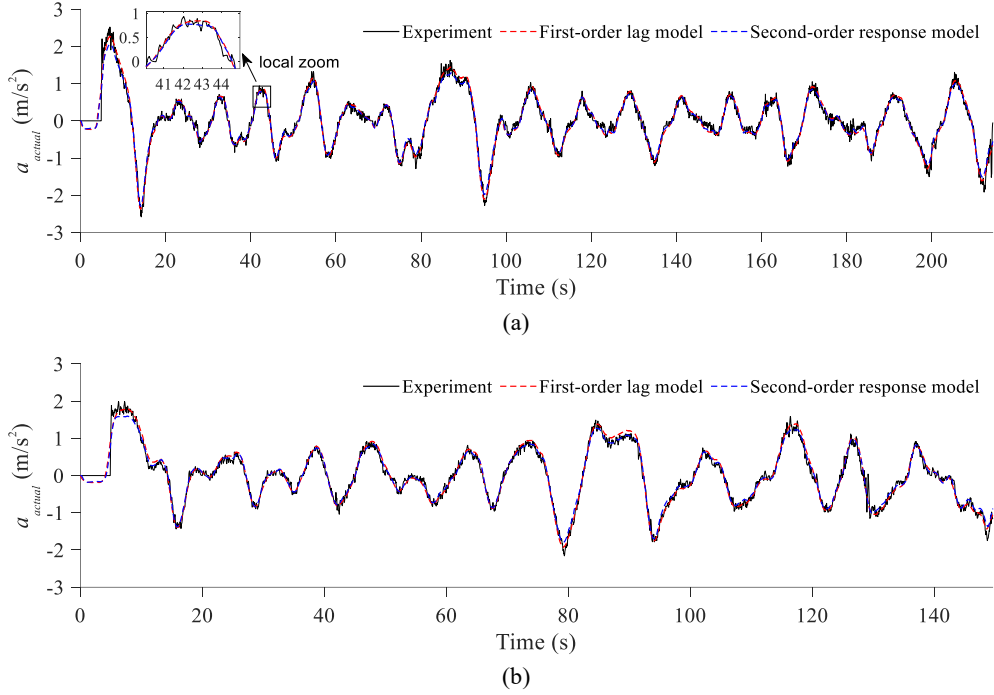

(a)

(b)

Fig. 2. Comparison of actual acceleration profile between experiment and simulation in (a) one calibration run and (b) one validation run of AV 6.

# 5. Experimental results

The experimental results of the 7-, 10-, 13-, 20-vehicle-platoon are presented in Section 5.1-5.4, respectively. Finally, the flow-density relation is measured and presented in Section 5.5.

## 5.1. Results of 7-vehicle-platoon

Firstly, we report the experimental results of 7-vehicle-platoon to reveal the characteristic of AV traffic. Figs. 3(a) and 3(b) present the velocity profile of each vehicle under the leading velocity of 20 km/h and 30 km/h, respectively. The overshooting phenomenon can be observed in the starting process. In the stationary state, the velocities fluctuate but the fluctuation amplitude is small. The corresponding speed spatiotemporal evolution diagrams are presented in Figs. 3(c) and 3(d), respectively.

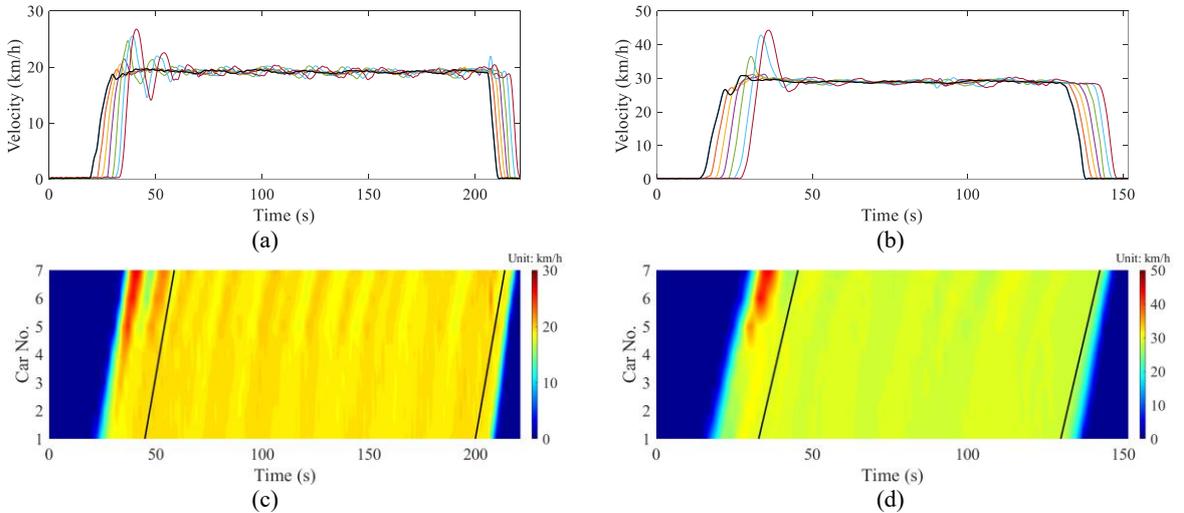

(a)

(b)

(c)

(d)

Fig. 3. (a, b) The velocity profiles and (c, d) the speed spatiotemporal evolution diagrams of 7-vehicle-platoon. (Left panel) $v_l$ = 20 km/h; (right panel) $v_l$ = 30 km/h. The concerned time intervals of stationary state are indicated by two oblique black lines in (c) an (d), considering the propagation of speed oscillation along the platoon.

To quantitatively capture the speed spatiotemporal evolution, the growth pattern of speed oscillation amplitude is studied. Over the concerned time interval of stationary state shown in Figs. 3(c) and 3(d), the standard deviation (STD) of speed for each individual vehicle is calculated. As shown in Fig. 4, the STDs of speed are very small and grow very slightly along the platoon.

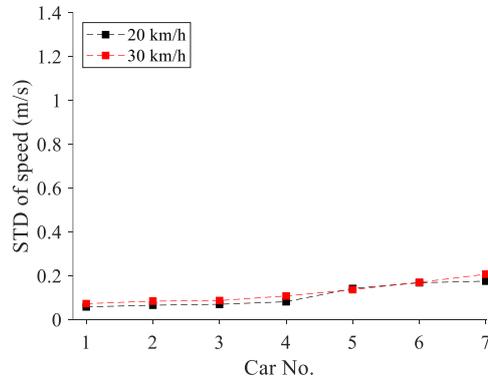

Fig. 4. The growth pattern of speed STD along the 7-vehicle-platoon.

Next, we investigate evolution of the average velocity of all cars and the platoon length. As shown in Fig. 5, both of them are rather stable in the stationary state. Under the leading velocity of 20 km/h and 30 km/h, the STD of average velocity is 0.1897 km/h and 0.2182 km/h, respectively; the STD of platoon length is 0.7848 m and 0.8347 m, respectively.

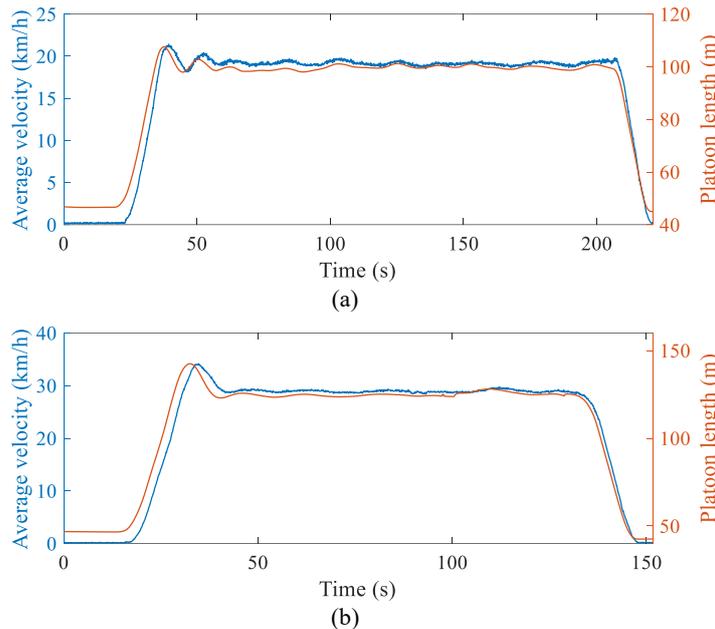

(a)

(b)

Fig. 5. Time series of average velocity and platoon length under (a) $v_l$ = 20 km/h; (b) $v_l$ = 30 km/h.

## 5.2. Results of 10-vehicle-platoon

Figs. 6(a) and 6(c) present the typical samples of speed spatiotemporal evolution diagrams of the benchmark in which all AVs switch into HV mode. One can see that vehicles in the rear part of the platoon experience remarkable oscillations. The stripe structure that exhibits formation and propagation of oscillations can be clearly observed. The typical samples of speed spatiotemporal evolution diagrams of the mixed platoon are presented in Figs. 6(b) and 6(d). Compared with that of the benchmark, the stripe structure is much blurred, which implies a more stable platoon.

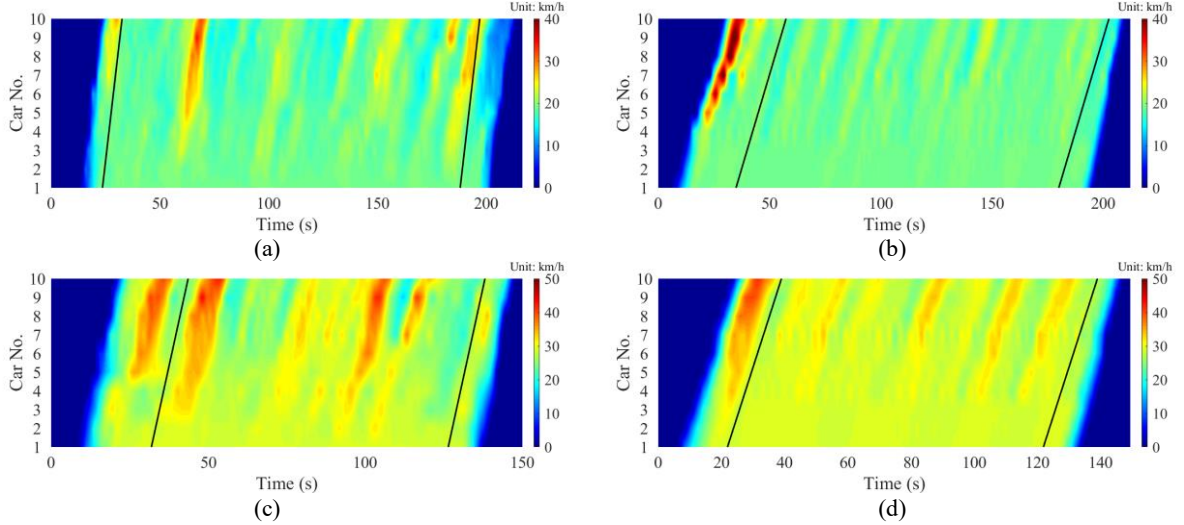

Fig. 6. The speed spatiotemporal evolution diagrams of 10-vehicle-platoon under (top panel) $v_l$ = 20 km/h; (bottom panel) $v_l$ = 30 km/h. (a) Run 1 and (c) Run 3 of the benchmark; (b) Run 1 and (d) Run 3 of the mixed platoon.

To make a quantitative comparison of the oscillation growth, the STDs of speed are calculated and compared between the benchmark and the mixed platoon in Fig. 7. One can see that the STDs of speed increase along the platoon, as reported in our previous studies (Huang et al., 2018; Jiang et al., 2014, 2015, 2018; Tian et al., 2016). The growth of speed STD in the mixed platoon is much slower than that in the benchmark, which indicates that a high MPR of AVs can significantly suppress the growth of speed oscillations.

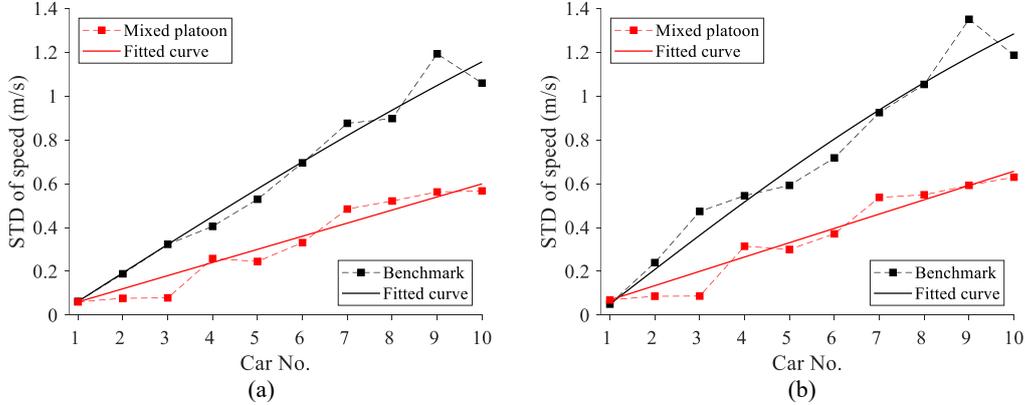

Fig. 7. Comparison of the growth pattern of speed STD along the 10-vehicle-platoon between the benchmark and the mixed platoon. (a) $v_l$ = 20 km/h; (b) $v_l$ = 30 km/h. The data of each curve are the average results of the Runs with the same settings.

Further, we compare the evolution of average velocity and platoon length in Fig. 8. It can be seen that the average velocity and platoon length in the benchmark fluctuate much greater than that in the mixed platoon. Table 10 compares the STDs of average velocity and platoon length. One can see that all the values in the benchmark are significantly larger. The STD of average velocity in the benchmark is more than 3 times that in the mixed platoon, and the STD of platoon length in the benchmark is 2-5 times that in the mixed platoon.

Finally, one can see that the length of the mixed platoon is much larger than that in the benchmark. The main reason is, the average spacing of AVs is larger in the AV mode than that in the HV mode. As shown later in Section 5.5, this leads to a lower traffic throughput in the mixed platoon.

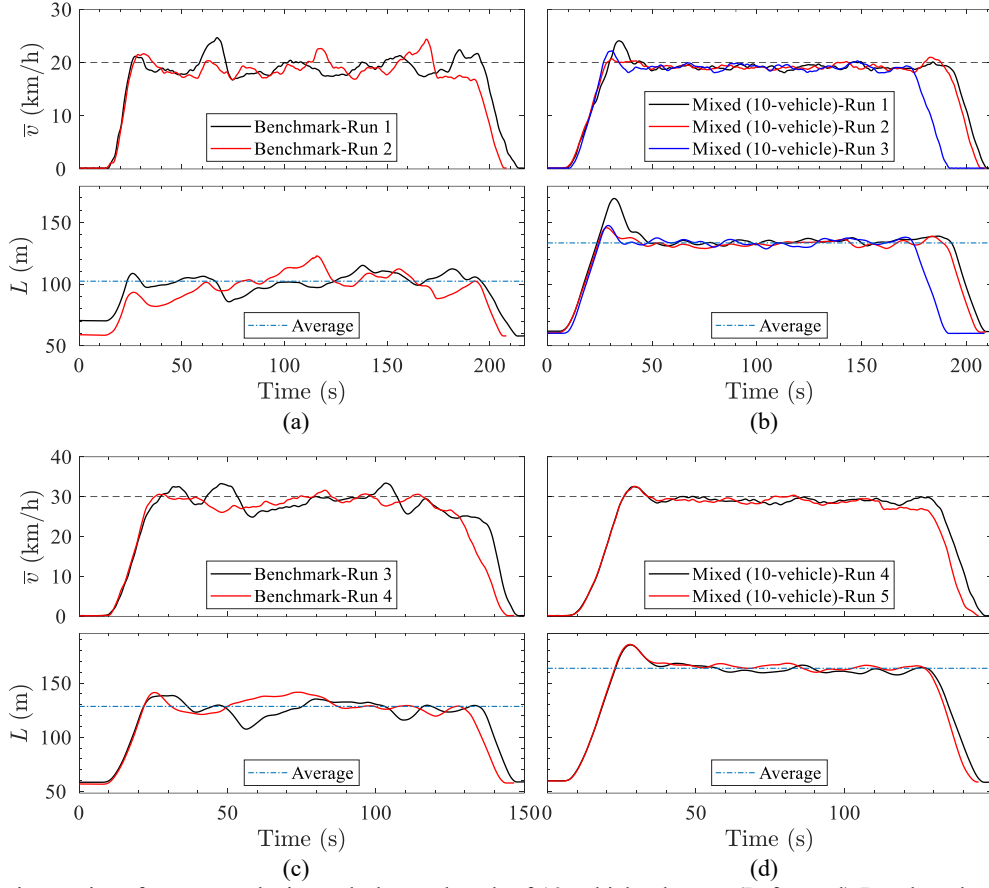

Fig. 8. The time series of average velocity and platoon length of 10-vehicle-platoon. (Left panel) Benchmark; (right panel) mixed platoon. (Top panel) $v_l = 20$ km/h; (bottom panel) $v_l = 30$ km/h. The blue chain line in each subfigure denotes the average platoon length over the concerned interval and the corresponding runs.

Table 10. The statistic result about average velocity and platoon length of 10-vehicle-platoon.

| $v_l$ | 20 km/h | | 30 km/h | |
|---|---|---|---|---|
| Exp. Type | Benchmark | Mixed platoon | Benchmark | Mixed platoon |
| $\sigma_{\bar{v}}$ (km/h) | 1.6195 | 0.4218 | 1.8013 | 0.5588 |
| $\sigma_L$ (m) | 7.8085 | 1.9031 | 6.3857 | 2.5545 |
| $\bar{L}$ (m) | 102.39 | 133.47 | 128.53 | 163.68 |

## 5.3. Results of 13-vehicle-platoon

The left and the right panels in Fig. 9 present the typical samples of speed spatiotemporal evolution diagrams of the benchmark and the mixed platoon, respectively. The diagrams are very similar to each other. Fig. 10 compares the STD of speed between the benchmark and the mixed platoon. The difference is not significant, which indicates that in the 13-vehicle-platoon, the six AVs do not have nontrivial impact on oscillation growth.

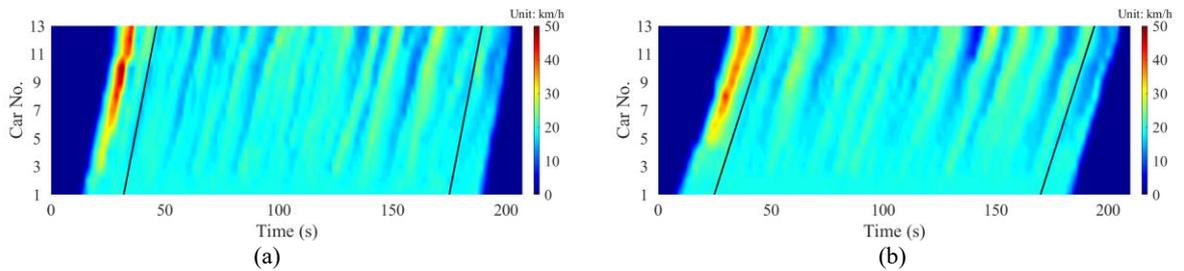

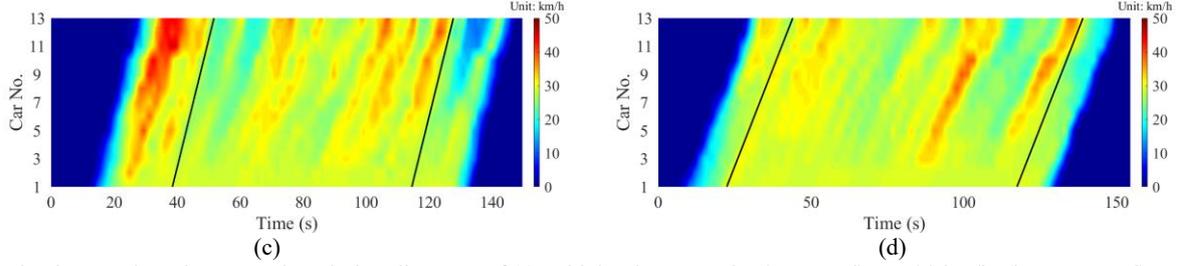

(c)                                                    (d)

Fig. 9. The speed spatiotemporal evolution diagrams of 13-vehicle-platoon under (top panel) $v_l = 20$ km/h; (bottom panel) $v_l = 30$ km/h. (a) Run 5 and (c) Run 7 of the benchmark; (b) Run 1 and (d) Run 3 of the mixed platoon.

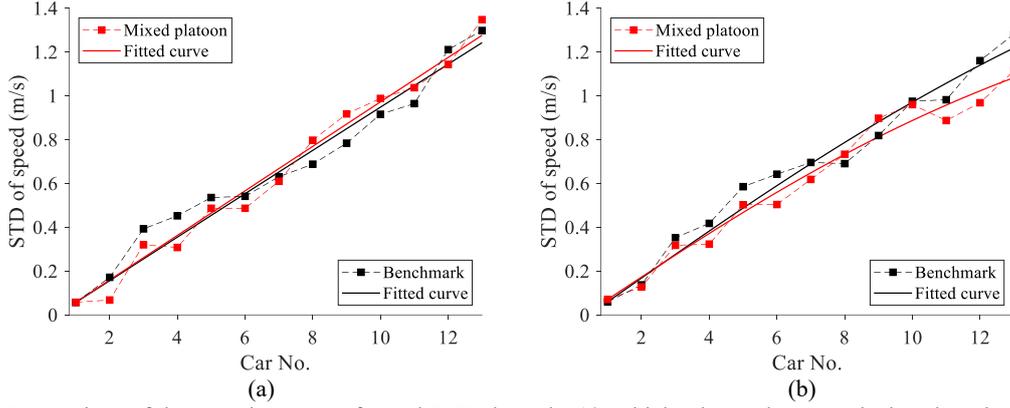

(a)                                                    (b)

Fig. 10. Comparison of the growth pattern of speed STD along the 13-vehicle-platoon between the benchmark and the mixed platoon. (a) $v_l = 20$ km/h; (b) $v_l = 30$ km/h.

The time series of average velocity and platoon length of the benchmark and the mixed platoon are presented in Fig. 11. Although the difference of speed STD becomes small, the difference of platoon length between the mixed platoon and the benchmark is still large, see also Table 11. The STD of average velocity and platoon length is presented in Table 11. One can see that the STD of average velocity and platoon length is roughly the same between mixed platoon and benchmark.

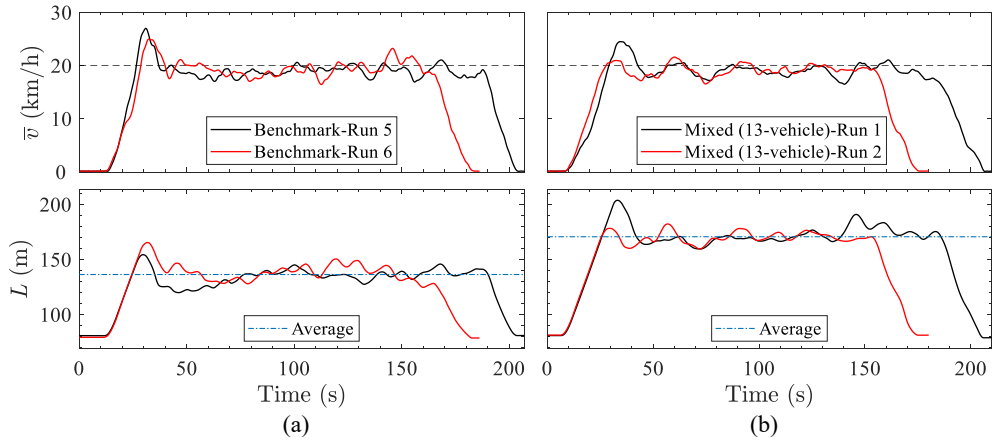

(a)                                                    (b)

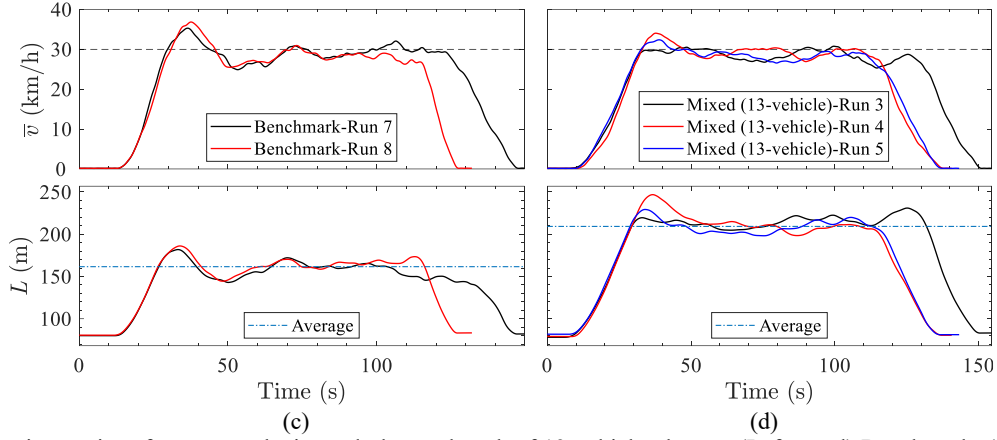

Fig. 11. The time series of average velocity and platoon length of 13-vehicle-platoon. (Left panel) Benchmark; (right panel) mixed platoon. (Top panel) $v_l$ = 20 km/h; (bottom panel) $v_l$ = 30 km/h.

Table 11. The statistic result about average velocity and platoon length of 13-vehicle-platoon.

| $v_l$ | 20 km/h | | 30 km/h | |
|---|---|---|---|---|
| Exp. Type | Benchmark | Mixed platoon | Benchmark | Mixed platoon |
| $\sigma_{\bar{v}}$ (km/h) | 1.0884 | 1.0390 | 1.4922 | 1.1233 |
| $\sigma_L$ (m) | 5.9908 | 5.8898 | 5.5351 | 5.3174 |
| $\bar{L}$ (m) | 136.42 | 170.73 | 161.59 | 209.27 |

## 5.4. Results of 20-vehicle-platoon

The left and the right panels in Fig. 12 present the typical samples of speed spatiotemporal evolution diagrams of the benchmark and the mixed platoon, respectively. Fig. 13 compares the STD of speed between the benchmark and the mixed platoon. As in the 13-vehicle-platoon experiment, the spatiotemporal diagrams are very similar to each other and the difference of speed STD is small, which indicates that in the 20-vehicle-platoon, the six AVs do not have significant impact on oscillation growth, either.

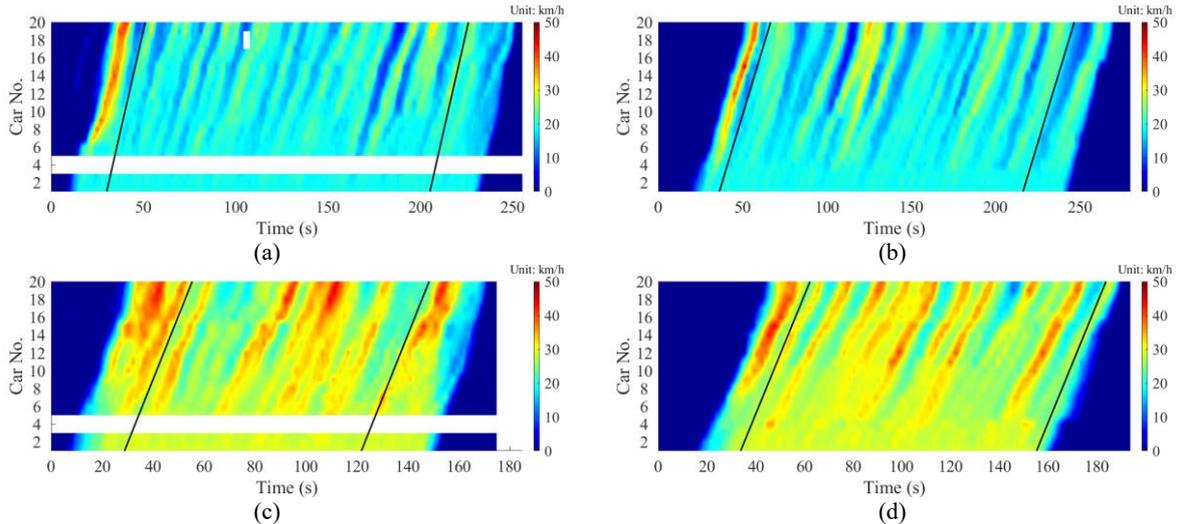

Fig. 12. The speed spatiotemporal evolution diagrams of 20-vehicle-platoon under (top panel) $v_l$ = 20 km/h; (bottom panel) $v_l$ = 30 km/h. (a) Run 9 and (c) Run 11 of the benchmark; (b) Run 1 and (d) Run 4 of the mixed platoon. The GPS signal of vehicle No.4 in Runs 9 and 11 is missing in the benchmark experiment.

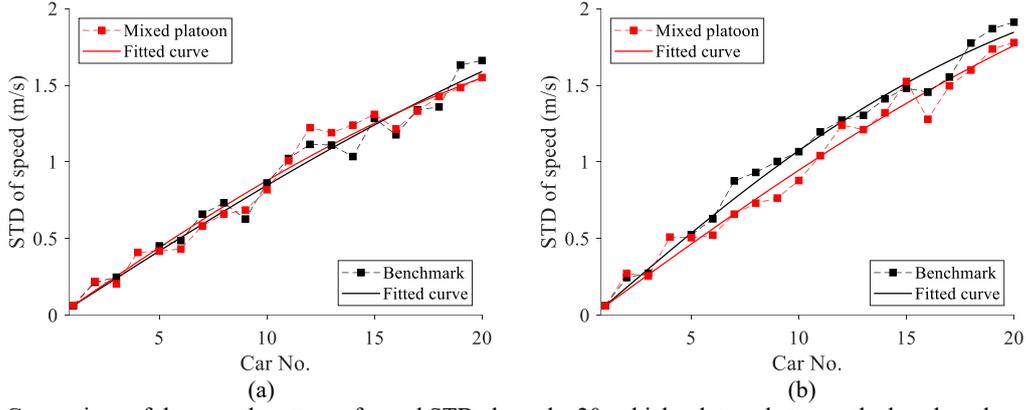

Fig. 13. Comparison of the growth pattern of speed STD along the 20-vehicle-platoon between the benchmark and the mixed platoon. (a) $v_l$ = 20 km/h; (b) $v_l$ = 30 km/h.

The time series of average velocity and platoon length of the benchmark and the mixed platoon are shown in Fig. 14. The average platoon length, the STD of average velocity and platoon length are presented in Table 12. It can be seen that the length of the mixed platoon is still remarkably larger than that of the benchmark, but the STD of average velocity and platoon length of the mixed platoon is close to that of the benchmark.

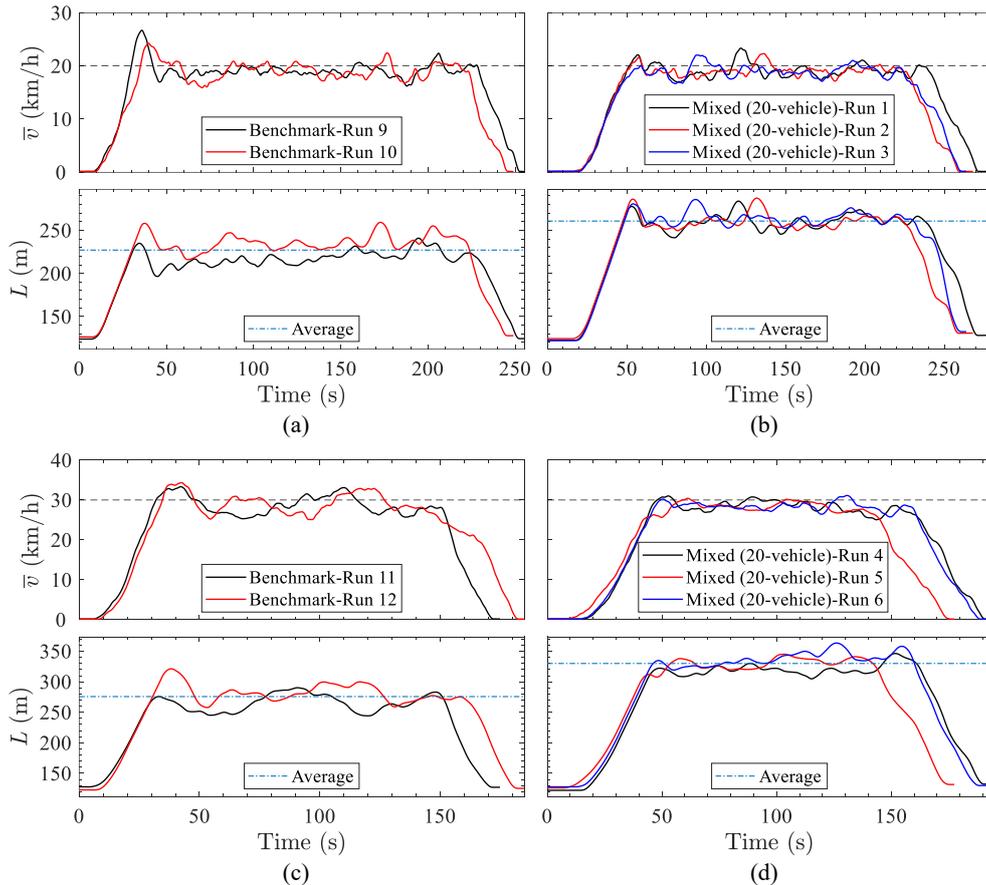

Fig. 14. The time series of average velocity and platoon length of 20-vehicle-platoon. (Left panel) Benchmark; (right panel) mixed platoon. (Top panel) $v_l$ = 20 km/h; (bottom panel) $v_l$ = 30 km/h.

Table 12. The statistic result about average velocity and platoon length of 20-vehicle-platoon.

| $v_l$ | 20 km/h | | 30 km/h | |
|---|---|---|---|---|
| Exp. Type | Benchmark | Mixed platoon | Benchmark | Mixed platoon |
| $\sigma_{\bar{v}}$ (km/h) | 1.1185 | 1.1021 | 2.3118 | 1.1247 |
| $\sigma_L$ (m) | 7.8369 | 7.8569 | 13.1166 | 8.9657 |
| $\bar{L}$ (m) | 227.09 | 260.74 | 275.73 | 329.94 |

## 5.5. Flow-density relation

Finally, the flow-density relation is presented in Fig. 15. It can be seen that given the leading velocity, the average density and the flow rate are the smallest in the 7-vehicle-platoon, and the largest in the benchmark. In the mixed platoon, the average vehicle density and the flow rate decrease with the increase of MPR of AVs. An exception is that under the leading velocity of 30 km/h, the average density and the flow rate are almost equal between 50% and 32% MPR. This might be due to stochasticity of HVs and limit runs of experiment. With more runs of experiment performed, the difference might become significant.

The results indicate a trade-off between traffic stability and throughput under the given setup of the six AVs, i.e., although the increase of MPR of AVs could possibly stabilize traffic flow, the flow rate is sacrificed. With the change of upper-level control algorithm (or tuning parameter value) of the six AVs, the trade-off might be erased, and both stability and throughput can be improved simultaneously. For related analysis, see Section 6.2.

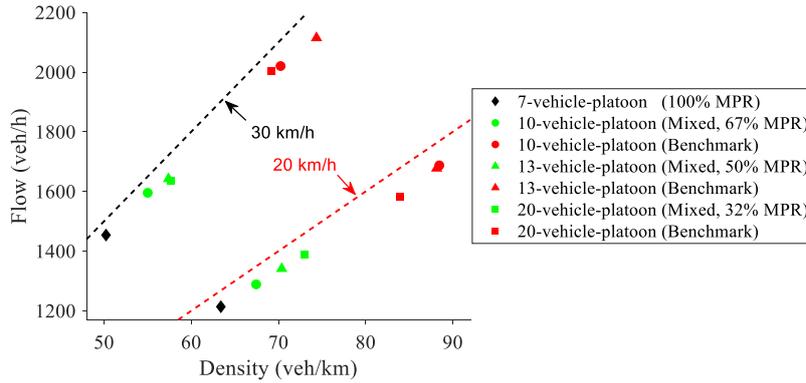

Fig. 15. The flow-density relation. The results are averaged over the Runs. The MPR is calculated excluding the leading vehicle.

# 6. Simulation study

This section firstly conducts a simulation study of the mixed traffic flow. Then sensitivity analysis of the parameters in the AV upper-level controller is performed.

## 6.1. Simulations of the mixed traffic flow

In the simulations, we use the two dimensional-intelligent driver model (2D-IDM) to describe the driving behaviors of HVs. The acceleration $a_n$ of vehicle $n$ at time $t$ is given by

$$a_n(t) = a_{\max}\left(1 - \left(\frac{v_n(t)}{v_{\max}}\right)^4 - \left(\frac{d_{n,desired}(t)}{d_n(t)}\right)^2\right) \qquad (6)$$

where $a_{\max}$ is the maximum acceleration; $v_n$ is the speed, and its maximum is $v_{\max}$; $d_{n,desired}$ is the desired spacing; the spacing $d_n(t) = x_{n-1}(t) - x_n(t) - L_{\text{veh}}$; $x_n$ is the position; $L_{\text{veh}}$ is the vehicle length.

The desired spacing depends on the current speed and its speed difference with the vehicle in front $n-1$, as follows.

$$d_{n,desired}(t) = d_0 + \max\left(v_n(t)T - \frac{v_n(t)\Delta v_n(t)}{2\sqrt{a_{\max}b}}, 0\right) \qquad (7)$$

where the velocity difference $\Delta v_n(t) = v_{n-1}(t) - v_n(t)$; $d_0$ is the safe stopping gap; $b$ is the comfortable deceleration; $T$ is the desired time gap.

Instead of a constant $T$ in the deterministic IDM (Treiber et al., 2000), $T$ changes stochastically between $T_{\min}$ and $T_{\max}$ in the 2D-IDM (Jiang et al., 2014; Xiong et al., 2019). The update process of $T$ in each simulation time step reads

$$T(t+\Delta t) = \begin{cases} \max\left\{T(t) - \Delta T, \tilde{T}(t+\Delta t)\right\}, & \text{if } \tilde{T}(t+\Delta t) < T(t) \\ \min\left\{T(t) + \Delta T, \tilde{T}(t+\Delta t)\right\}, & \text{otherwise} \end{cases} \qquad (8)$$

where $\Delta T$ is the change limitation; $\tilde{T}(t)$ is the tentative desired time gap that changes stochastically as follows.

$$\tilde{T}(t+\Delta t) = \begin{cases} T_{\min} + r \cdot (T_{\max} - T_{\min}), & \text{with probability } p \\ \tilde{T}(t), & \text{otherwise} \end{cases} \qquad (9)$$

where $r \in [0,1]$ is a uniformly distributed random number; $\Delta t$ is the simulation time step and is set to 0.1 s.

The calibration is conducted using the genetic algorithm with the goodness-of-fit function to minimize the sum of relative error between the measurements of performance (MoPs) of simulation and experiment with respect to the STDs of speed, platoon length, STDs of platoon length and average platoon velocity, which reads

$$\min \sum_{Run} \frac{1}{M} \cdot \sum_{k=1}^{M} \left[ \frac{1}{N} \cdot \sum_{j=1}^{N} \left| 1 - \frac{\sigma_{v,j,k,Run}^{\text{sim}}}{\sigma_{v,j,Run}^{\text{exp}}} \right| + \frac{1}{T} \sum_{t=1}^{T} \left| 1 - \frac{L_{t,k,Run}^{\text{sim}}}{L_{t,Run}^{\text{exp}}} \right| + \left| 1 - \frac{\sigma_{L,k,Run}^{\text{sim}}}{\sigma_{L,Run}^{\text{exp}}} \right| + \left| 1 - \frac{\sigma_{\bar{v},k,Run}^{\text{sim}}}{\sigma_{\bar{v},Run}^{\text{exp}}} \right| \right] \qquad (10)$$

where $M$ is the number of repeated simulations; $N$ is the number of vehicles in the Run; $T$ is the time length; $\sigma_{v,j,k,Run}^{\text{sim}}$ and $\sigma_{v,j,Run}^{\text{exp}}$ denote the STD of speed of vehicle $j$ in the $k$th simulation and in the experimental run, respectively. $L_{t,k,Run}^{\text{sim}}$ and $L_{t,Run}^{\text{exp}}$ denote the platoon length at time $t$ in the $k$th simulation and in the experimental run, respectively; $\sigma_{L,k,Run}^{\text{sim}}$ and $\sigma_{L,Run}^{\text{exp}}$ denote the STD of platoon length in the $k$th simulation and in the experimental run, respectively; $\sigma_{\bar{v},k,Run}^{\text{sim}}$ and $\sigma_{\bar{v},Run}^{\text{exp}}$ denote the STD of average platoon velocity in the $k$th simulation and in the experimental run, respectively.

The experimental Runs of all benchmarks and the 13-vehicle mixed platoon are used for calibration, while the Runs of the 10- and 20-vehicle mixed platoon are used for validation. We set $v_{\max} = 80$ km/h and $L_{\text{veh}} = 5$ m in the simulations. The calibrated model parameter values are shown in Table 13.

Table 13. Calibrated parameter values.

| Parameter | Unit | Lower bound | Upper bound | Calibrated |
|-----------|------|-------------|-------------|------------|
| $a_{\max}$ | m/s² | 0.5 | 3 | 1.1254 |
| $b$ | m/s² | 0.5 | 6 | 5.5678 |
| $T_{\min}$ | s | 0 | 2 | 0.3049 |
| $T_{\max}$ | s | 1.5 | 5 | 1.5532 |
| $\Delta T$ | s | 0.01 | 0.1 | 0.0218 |
| $p$ | — | 0 | 1 | 0.3268 |
| $d_0$ | m | 0.5 | 4 | 1.5255 |

Simulations demonstrate that the 2D-IDM together with the AV control algorithm and controller model well reproduces the dynamics of the mixed traffic flow. Instead of enumerating all the simulation results, we show a few typical ones. Fig. 16 compares the growth pattern of speed STD between simulation and experiment, which shows that the speed oscillation growth can be quantitatively well reproduced. Fig. 17 compares the time series of average velocity and platoon length between simulation and experiment. It can be seen that the simulated platoon length is close to the experimental one, and the fluctuations of average velocity and platoon length in the simulation are very similar to those in the experiment. The average platoon length and the STD of average velocity and platoon length of the simulated mixed traffic flow are reported in Table 14.

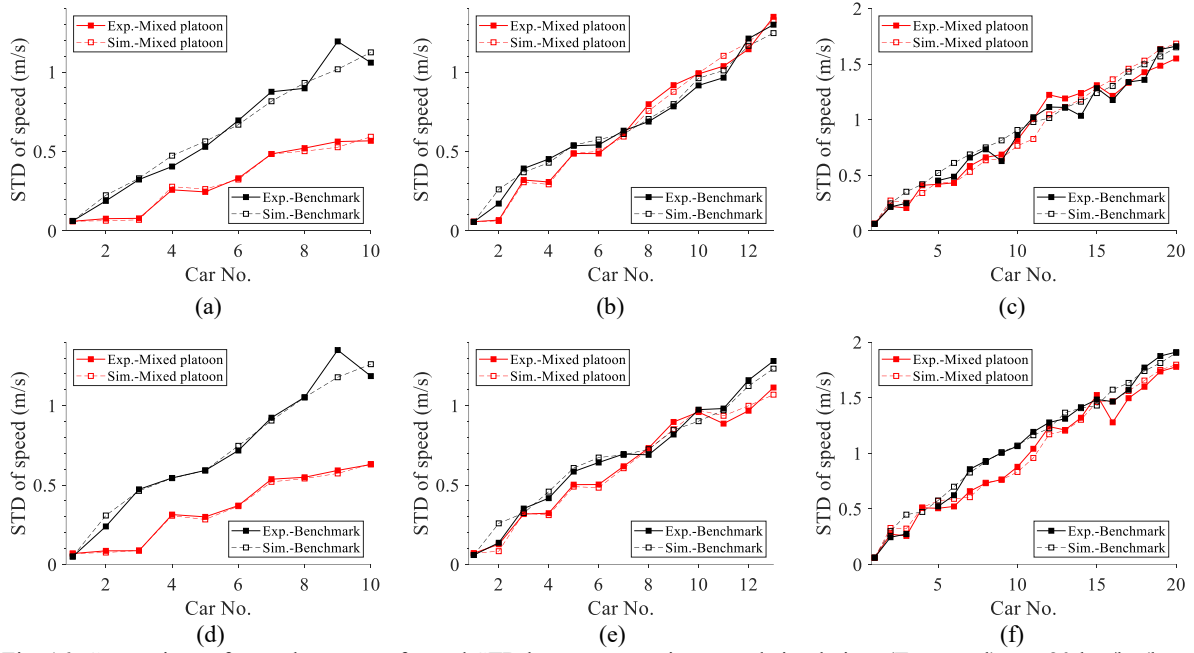

Fig. 16. Comparison of growth pattern of speed STD between experiment and simulation. (Top panel) $v_l$ = 20 km/h; (bottom panel) $v_l$ = 30 km/h. (Left panel) 10-vehicle-platoon; (middle panel) 13-vehicle-platoon; (right panel) 20-vehicle-platoon.

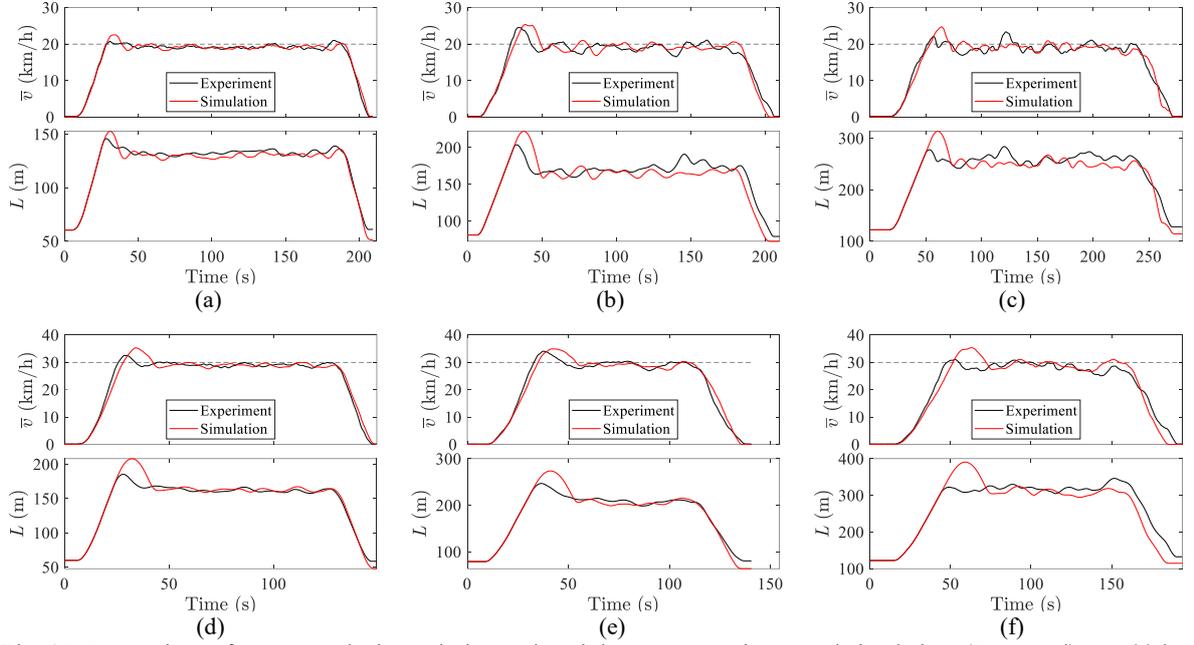

Fig. 17. Comparison of average velocity and platoon length between experiment and simulation. (Top panel) $v_l$ = 20 km/h; (bottom panel) $v_l$ = 30 km/h. (Left panel) 10-vehicle-platoon: (a) Run 2, (d) Run 4; (middle panel) 13-vehicle-platoon: (b) Run 1, (e) Run 4; (right panel) 20-vehicle-platoon: (c) Run 1, (f) Run 4.

Table 14. The statistic result about simulated average velocity and platoon length. The experiment one is presented in the bracket as a reference.

| $v_l$ | 20 km/h | | | 30 km/h | | |
|---|---|---|---|---|---|---|
| Platoon size | 10-vehicle | 13-vehicle | 20-vehicle | 10-vehicle | 13-vehicle | 20-vehicle |
| $\sigma_v$ (km/h) | 0.4563 (0.4218) | 0.8013 (1.0390) | 0.7159 (1.1021) | 0.4778 (0.5588) | 0.9076 (1.1233) | 1.1242 (1.1247) |
| $\sigma_L$ (m) | 1.8133 (1.9031) | 4.1785 (5.8898) | 5.4699 (7.8569) | 1.7299 (2.5545) | 4.3984 (5.3174) | 7.3495 (8.9657) |
| $\bar{L}$ (m) | 129.97 (133.47) | 165.33 (170.73) | 250.24 (260.74) | 162.72 (163.68) | 205.64 (209.27) | 308.15 (329.94) |

Furthermore, the simulated flow-density relation is measured and presented in Fig. 18. The phenomena, (i) the flow rate of the 7-vehicle-platoon is the smallest; (ii) the flow rate of the benchmarks is the largest; and (iii) both flow rate and density decrease with the increase of MPR of AVs, can be well captured.

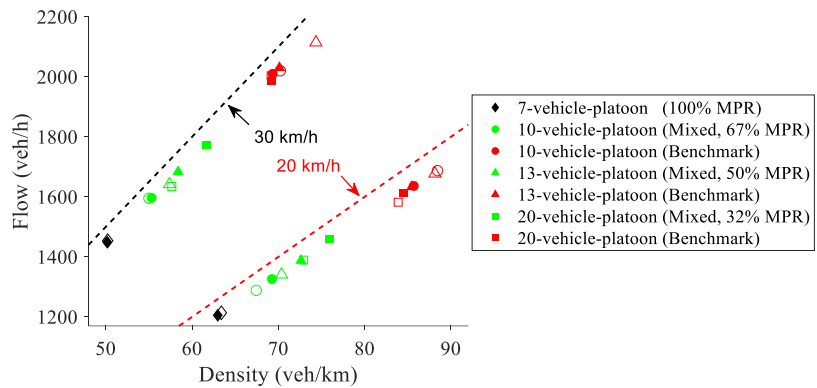

Fig. 18. Comparison of flow-density relation between experiment and simulation. The filled symbols are simulation results and the unfilled symbols are the corresponding experimental ones.

## 6.2. Sensitivity analysis

In the future work, we plan to carry out more experiments using the six AVs. To guide the experiment design, this subsection performs sensitivity analysis of the upper-level control parameters of the AVs. It is known that the lower-level controller of AVs is determined by vehicle mechanical performance. Given the vehicle, the lower-level controller as well as the parameters is usually not changeable. Therefore, we still use the lower-level controller model and parameters for the six AVs. However, the upper-level control algorithm as well as the parameters can be arbitrarily set by users, which would impact the performance of AVs and thus the mixed traffic flow. Herein, we still use the constant time gap algorithm and only tune the parameters. We mainly present the results in a 20-vehicle-platoon with homogeneous distribution of six AVs as in the experiment.

Firstly, the impact of control gains $k_v$ and $k_g$ on the oscillation growth is studied. As shown in Fig. 19(a), with the increase/decrease of $k_v$, the speed oscillation grows slower/faster, which indicates that the platoon stability is enhanced/weakened. Moreover, with the decrease of $k_v$, collision would occur in the starting process, as a result of the overshooting. For example, at $k_v = 0.2$ s$^{-1}$, the 15th vehicle would collide into the 14th vehicle, see Fig. 19(b).

The impact of $k_g$ on the oscillation growth is different. There exists a critical value $k_g \approx 0.25$ s$^{-2}$, under which speed oscillation grows the fastest. With either the increase or the decrease of $k_g$, the platoon becomes more stable, see Fig. 20. Nonetheless, for $k_g > 0.25$ s$^{-2}$, the initial growth of oscillation in the front part of 8 vehicles is almost independent of $k_g$.

Next, we investigate the effect of $T_g$. As shown in Fig. 21, the effect of $T_g$ on the oscillation growth is like that of $k_v$. Platoon stability enhances/weakens with the increase/decrease of $T_g$.

Now, we study the impact of $k_v$, $k_g$, and $T_g$ on platoon length. As shown in Fig. 21(a), since the platoon becomes more stable, the platoon length slightly decreases from 252.5 m to 247.4 m when $k_v$ increases from 0.3 to 1 s$^{-1}$. On the other hand, since the spacing increases linearly with $T_g$, the platoon length significantly increases from 251 to 281.6 m, when $T_g$ increases from 1.4 to 2.6 s, see Fig. 21(c). Finally, as shown in Fig. 21(b), the platoon length changes non-monotonically with the increase of $k_g$. This is also related to the platoon stability. The more stable the platoon is, the smaller the length is.

We have also tested the following scenarios: (i) the 7- and 13-vehicle-platoon; (ii) the platoon with more than 20 vehicles, in which MPR of AVs is smaller than 32% because we have only six AVs; (iii) AVs are distributed randomly in the platoon. Similar results are observed as reported above. In our future work, we will carry out experiments to examine the results of sensitivity analysis.

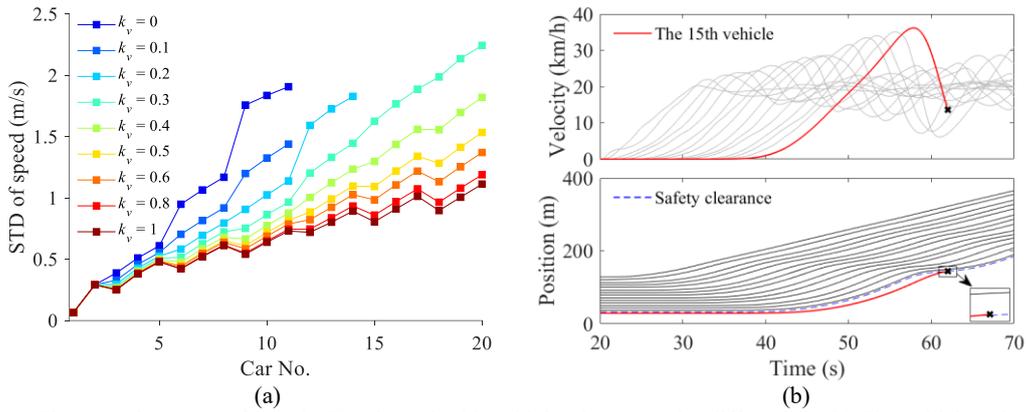

Fig. 19. (a) The growth pattern of speed STD along the 20-vehicle-platoon under different $k_v$. (b) The collision situation at $k_v$ = 0.2 s$^{-1}$. The short platoon is due to that collision would occur if platoon length increases.

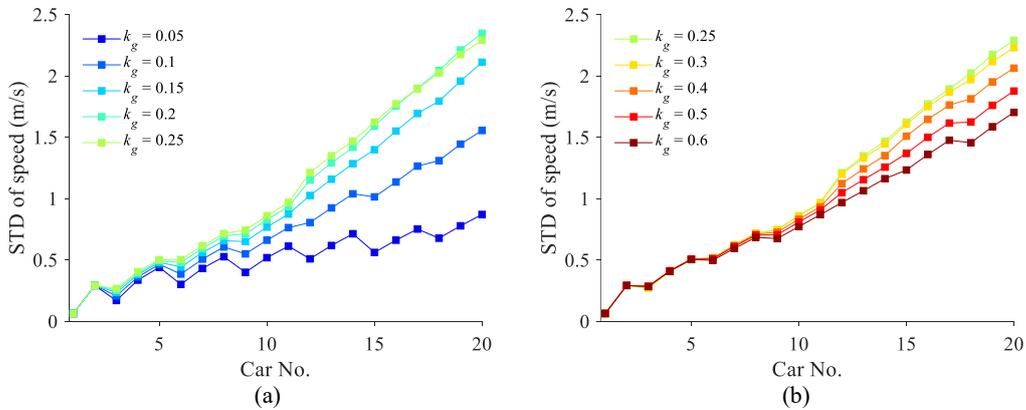

Fig. 20. The growth pattern of speed STD along the 20-vehicle-platoon under different $k_g$.

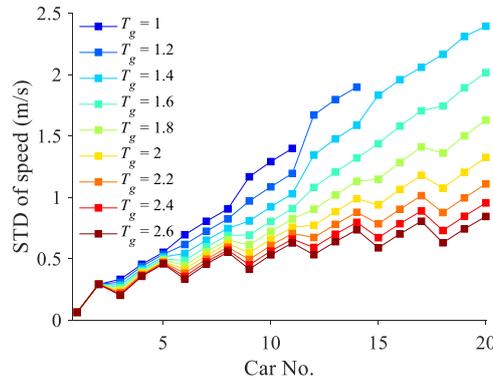

Fig. 21. The growth pattern of speed STD along the 20-vehicle-platoon under different $T_g$.

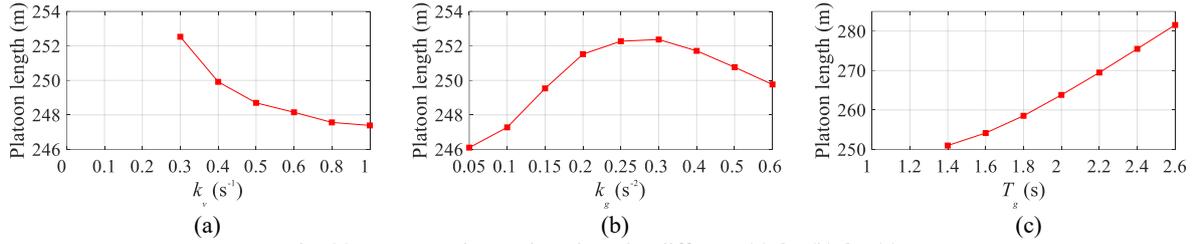

Fig. 22. Average platoon length under different (a) $k_v$; (b) $k_g$; (c) $T_g$.

# 7. Conclusion

Emerging AV technologies are expected to have profound impacts on traffic system. However, there will be a long transition before the coming of 100% deployment (Mahmassani, 2016). In the past several decades, many efforts have been devoted to mixed traffic flow of AVs and HVs. While there is still no consensus on the impact of AVs, the field experiments with commercial AVs have demonstrated the important role of field experiments to understand traffic dynamics of AVs (Li et al., 2021).

Recent experimental study of car following behavior of HVs demonstrates that traffic oscillation grows in a concave way along the platoon (Jiang et al., 2014, 2015, 2018). The finding reveals the mechanism of traffic instability and exhibits how traffic flow develops into jams. Nevertheless, the formation and growth of traffic oscillation in mixed traffic flow of AVs and HVs have not been studied, which is expected to play an important role in understanding mixed traffic dynamics.

Motivated by the fact, we carried out a mixed platoon experiment on a straight track with six developable AVs that are distributed homogeneously in the platoon. The constant time gap car-following policy was adopted for the AVs and the gap was set to 1.5 s. In the experiment, the leading vehicle moves with constant speed, and we investigate the growth of oscillations in platoon of different size.

The experiment shows that in the 7-vehicle-platoon, the oscillations grow only slightly. In the 10-vehicle-platoon, the AVs could still significantly suppress the growth of oscillations. Nevertheless, the AVs do not have significant impact on oscillation growth in the 13- and 20-vehicle-platoon. On the other hand, with the decrease of MPR of AVs, average density of the vehicles and flow rate of the platoon increase, which demonstrates a trade-off between traffic stability and throughput.

We also carry out the simulation study on the oscillation growth in the mixed platoon. Simulations demonstrate that the 2D-IDM together with the AV control algorithm and controller model well reproduces the dynamics of the mixed traffic flow. Finally, we perform sensitivity analysis to guide the future experiment design. It is shown that by tuning the parameters, both stability and throughput can be improved simultaneously. In our future work, experiments will be performed to examine the sensitivity analysis results.

# Acknowledgments

This work was supported in part by the Fundamental Research Funds for the Central Universities under Grant 2021JBZ107 and the National Natural Science Foundation of China under Grant No.71621001 and 71931002.

# Appendix A. The detail information about vehicles and drivers.

Table A.1. The detail information about vehicles and drivers. Each vehicle has a unique driver throughout the experiment. All vehicles are automatic transmission ones. √ denotes that the vehicle was present for a given experimental Run.

| Vehicle No. | Make and model | Vehicle type | Platoon size (vehicles) | | | | $L \times W \times H$ (mm×mm×mm) | Swept volume (L) | Driving experience (years) |
|---|---|---|---|---|---|---|---|---|---|
| | | | 7 | 10 | 13 | 20 | | | |
| 1 | VOLKSWAGEN PASSAT | Equipped with CC system | √ | √ | √ | √ | 4870×1834×1472 | 1.8T | 7 |
| 2 | KIA K5 | HV | | √ | | √ | 4855×1835×1475 | 2.0 | 17 |
| 3 | HONGQI H5 | | | | √ | √ | 4945×1845×1470 | 1.8T | 12 |
| 4 | MAZDA 3 | | | | | √ | 4532×1755×1465 | 1.6 | 26 |
| 5 | BUICK GL8 | | | | √ | √ | 5213×1847×1750 | 2.4 | 20 |
| 6 | VOLKSWAGEN SAGITAR | | | | | √ | 4753×1800×1462 | 1.4T | 10 |
| 7 | HONDA ACCORD | | | | √ | √ | 4893×1862×1449 | 2.0 | 28 |
| 8 | NISSAN SUNNY | | | √ | | √ | 4456×1696×1514 | 1.5 | 9 |
| 9 | VOLKSWAGEN SAGITAR | | | | √ | √ | 4753×1800×1462 | 1.2T | 13 |
| 10 | BEIJING-HYUNDAI ELANTRA | | | | | √ | 4610×1800×1450 | 1.5 | 7 |
| 11 | BUICK GL8 | | | | √ | √ | 5213×1847×1750 | 2.4 | 13 |
| 12 | VOLKSWAGEN BORA | | | | | √ | 4663×1815×1462 | 1.5 | 14 |
| 13 | BUICK GL8 | | | | √ | √ | 5213×1847×1750 | 2.5 | 27 |
| 14 | KIA K3 | | | | | √ | 4600×1780×1445 | 1.6 | 14 |
| 15 | NISSAN SYLPHY | | | √ | | | 4631×1760×1503 | 1.6 | 18 |
| 16, AV 1 | CHANGAN AUTO CS55 E-Rocks | AV | √ | √ | √ | √ | 4515×1860×1690 | EV | 7 |
| 17, AV 2 | CHANGAN AUTO CS55 E-Rocks | | √ | √ | √ | √ | 4515×1860×1690 | EV | 9 |
| 18, AV 3 | CHANGAN AUTO CS55 E-Rocks | | √ | √ | √ | √ | 4515×1860×1690 | EV | 4 |
| 19, AV 4 | CHANGAN AUTO CS55 PLUS | | √ | √ | √ | √ | 4515×1865×1680 | 1.5 | 8 |
| 20, AV 5 | CHANGAN AUTO CS55 E-Rocks | | √ | √ | √ | √ | 4515×1860×1690 | EV | 11 |
| 21, AV 6 | BAIC MOTOR ARCFOX αT | | √ | √ | √ | √ | 4788×1940×1683 | EV | 11 |

* Electric Vehicle is abbreviated as EV.